\documentclass[slac_one]{revtex4}
\usepackage{subfigure}
\usepackage{graphicx}
\usepackage{fancyhdr}
\newcommand{\bbar}{$b\overline{b}$~}

\begin{document}

\title{{\small{Hadron Collider Physics Symposium (HCP2008),
Galena, Illinois, USA}}\\ 
\vspace{12pt}
B Physics at LHC} 

%

\author{A. Sarti}
\affiliation{Laboratori Nazionali di Frascati, Via E.~Fermi 40, 00044, Frascati, Italy}
%

\begin{abstract}
Three experiments, among the LHC project, are getting ready to explore
the b quark flavour sector.
While ATLAS and CMS are general purpose experiments,
where the study of B mesons is going to proceed in parallel with 
the Higgs boson and supersymmetry searches, the
LHCb experiment is devoted to B physics studies. 
The key parameters entering the physics analyses and the
performances achieved in all the three experiments
are presented.
Given the large B physics program foreseen in the
LHC experiments, the studies reported in this paper have
been selected as those with higher likelihood
to provide solid and interesting new
results on Standard Model validation and
New Physics processes search with early data.
\end{abstract}

\maketitle

\thispagestyle{fancy}


\section{INTRODUCTION} 

In the context of the large effort performed within
the LHC~\cite{lhc:det} experiments in order to finalize the construction
of the Standard Model (SM) and to shed some light on the
building blocks of its future extension, the B physics
program plays a very important role.

In a complementary way to the direct searches
of Higgs boson(s) and new particles, the New Physics (NP)
processes can be studied by making precision
measurements in the flavour sector, where we do expect
significant effects if the NP scale is not completely decoupled
from the TeV energy range.

In this framework, the B meson system is a natural place where
to look. The $B_d$ mesons properties have already been studied in detail
by the B-factories experiments (CLEO, BaBar, Belle) yielding and
impressive agreement with the SM expectations. 
The properties of the $B_s$ mesons,
not available at e$^+$e$^-$ colliders\footnote{The Belle experiment, 
since last year, started a physics program at 
the $\Upsilon$(5S). The
small statistics of $B_s$ mesons collected, 
will mainly be used to measure some
absolute branching ratios.}, are much less constrained and there's
still room to account for significant contributions from NP effects. 
Moreover, the study of the B system with high statistics samples
of different b hadrons can be used to further constrain the
SM CP violation mechanism described by the 
Cabibbo-Kobayashi-Maskawa (CKM) matrix.

Three LHC experiments, LHCb~\cite{lhcb:det}
ATLAS~\cite{atlas:det} and CMS~\cite{cms:det},
have defined a research program that includes the search of 
rare decays and NP signatures
trough precision measurements of the meson decays.

While LHCb~\cite{lhcb:det} is devoted and optimized
for the measurement of the B meson properties,
the other two are general purpose experiments
where the study of B mesons proceeds in parallel with 
the Higgs boson and supersymmetry searches.

The doability of such B physics programs at the hadron machines
is demonstrated by the recent results of Tevatron experiments
(CDF, D0 \cite{tev:det}) 
with their nice measurements of $\Delta m_s$~\cite{tev:dms} and
$\Delta\Gamma_s$ and $\phi_s$~\cite{tev:bs}. 

The LHC environment, the expected resolutions of the various detectors
and their triggering strategies 
are reviewed in section \ref{sec:lhc},
while section \ref{sec:bprog} contains the detail description of
the physics program and of the expected performances 
that can be achieved with the data samples collected
during the first year of operation of LHC.

\section{LHC DETECTORS FOR B PHYSICS}
\label{sec:lhc}

The parameters summarizing the LHC environment
for B physics studies (center of mass energy, 
cross sections, luminosity)
are reported in Table~\ref{tab:kpar}
for the LHCb, ATLAS and CMS experiments.

The very large cross section for \bbar production,
yielding $O$(10$^4$) \bbar pairs per second 
(L = 10$^{32}$~cm$^{-2}$ s$^{-1}$) with all the b-hadrons species
being produced, extend down to 10$^{-9}$
the Branching Ratio (BR) measurement capability 
for the rare B decays.

For this aim a huge effort is 
needed in order to reduce the large background coming 
from the pp interactions, whose production cross section 
is two orders of magnitude higher than the b quarks production one.

Attention should also be paid to the choice of running luminosity (L):
while larger L provides larger data samples,
the number of interactions
per crossing increase rapidly with L, creating problems to the
B hadron reconstruction and identification (see Table~\ref{tab:kpar}).

\begin{figure*}[t]
\centering
\includegraphics[width=70mm]{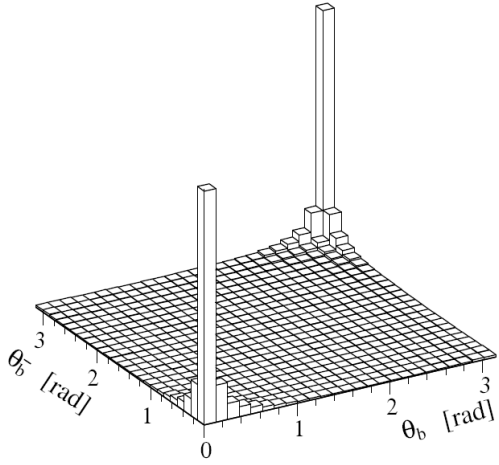}
\caption{Production angle of B vs. angle of B
in the laboratory (in units of rad.), calculated using
PYTHIA. The peaks in the forward directions shows
the correlation between their respective production
directions.\label{fig:bbcorr}} 
\end{figure*}

The main difference in the visible cross section
between the LHCb detector, a single arm
forward spectrometer accepting events 
with 1.8$<|\eta|<$5, and the ATLAS/CMS detectors, 
that are accepting events with $|\eta|<$2.5,
comes from the lower p$_T$
threshold that can be achieved in the former.
The production of \bbar pairs in LHC, according to 
PYTHIA simulations, occurs with the \bbar pairs highly collimated
(as shown in Figure \ref{fig:bbcorr}): this also favor the 
LHCb forward design with respect to the 4$\pi$ design of the 
others two experiments.

\begin{table}[t]
\begin{center}
\caption{Center of mass energy, cross sections, luminosity and number of interactions per crossing expected in the LHCb, ATLAS and CMS experiments.}
\begin{tabular}{|l|c|c|}
\hline
\textbf{Parameter} & \textbf{LHCb} & \textbf{ATLAS,CMS} \\ \hline
Energy (TeV) & 14 & 14\\
$\sigma$(pp) (mb) & 100 & 100 \\
$\sigma$(\bbar) $\mu$b & 500 & 500 \\
Visible $\sigma$(\bbar) $\mu$b & 230 & 100 \\
L (cm$^{-2}$ s$^{-1}$) & 2$\cdot$10$^{32}$ & 1$\cdot$10$^{34}$ \\
\#$_{int}$/crossing  & 1 & 23\\
\hline
\end{tabular}
\label{tab:kpar}
\end{center}
\end{table}

The main experimental ingredients entering the B mesons decays study,
like the tracking, particle identification and tagging performances
and proper time measurements,
are presented in detail in the next paragraphs, as well as the
currently implemented trigger strategies.

\subsection{Tracking}
\label{sec:expAM}

The track reconstruction performance is of primary 
importance in the various
steps of the B meson decays analysis: the capability of triggering
interesting events, the event selection and background suppression.

In the ATLAS and CMS detectors a large part of the B physics 
program is performed using channels involving muons, either as
the particles triggering the acquisition (high p$_T$ tracks)
or in the final states ($J/\psi$ decays).
The high overall muon reconstruction efficiency ($>$98\%, CMS) 
achieved is due to the 
good muon transverse momentum resolution (1$-$2\%)
and impact parameter resolution (10~$\mu$m) 
in the trackers.

The LHCb detector is able to use kaons and pions in the trigger, in addition
to the electrons and muons that are available also in ATLAS and CMS. 
The expected performances are: a relative momentum resolution
in the range 0.3$-$0.5\% depending on the track momentum and
a very high reconstruction efficiency ($>$95\%) for particles travelling
in the whole detector with a small ($<$4\%) contamination
of ghost tracks (having applied a p$_T >$ 0.5~GeV/c cut).

The track impact parameter resolution is $\sim$30~$\mu$m and the typical
B meson mass resolution, obtained in some benchmark channels studying the
reconstruction of fully simulated events, 
are given in Table~\ref{tab:LBphyspar}.
Those values are remarkable when considering the very high
charged particles multiplicity environment of the experiment:
$\sim$70 charged particles are expected for each \bbar event
at nominal luminosity (see Table \ref{tab:kpar}).

\begin{table}[t]
\begin{center}
\caption{Typical B mesons mass and proper time 
resolutions in the LHCb experiment. The last raw contains
the value obtained for the $B_s$ meson mass resolution 
after having applied a mass constraint on
the $J/\psi$.}
\begin{tabular}{|l|c|}
\hline
\textbf{Parameter} & \textbf{Value} \\ \hline
Proper time & 40 fs \\
Mass ($B_s \to \mu\mu$) & 18 MeV/c${^2}$ \\
Mass ($B_s \to D_s\pi$) & 14 MeV/c${^2}$ \\
Mass ($B_s \to J/\psi\phi$) & 14 MeV/c${^2}$ \\
Mass ($B_s \to J/\psi\phi$) & 8 MeV/c${^2}$ \\
\hline
\end{tabular}
\label{tab:LBphyspar}
\end{center}
\end{table}

\subsection{Particle Identification}

\begin{figure*}[t]
\centering
\includegraphics[width=70mm]{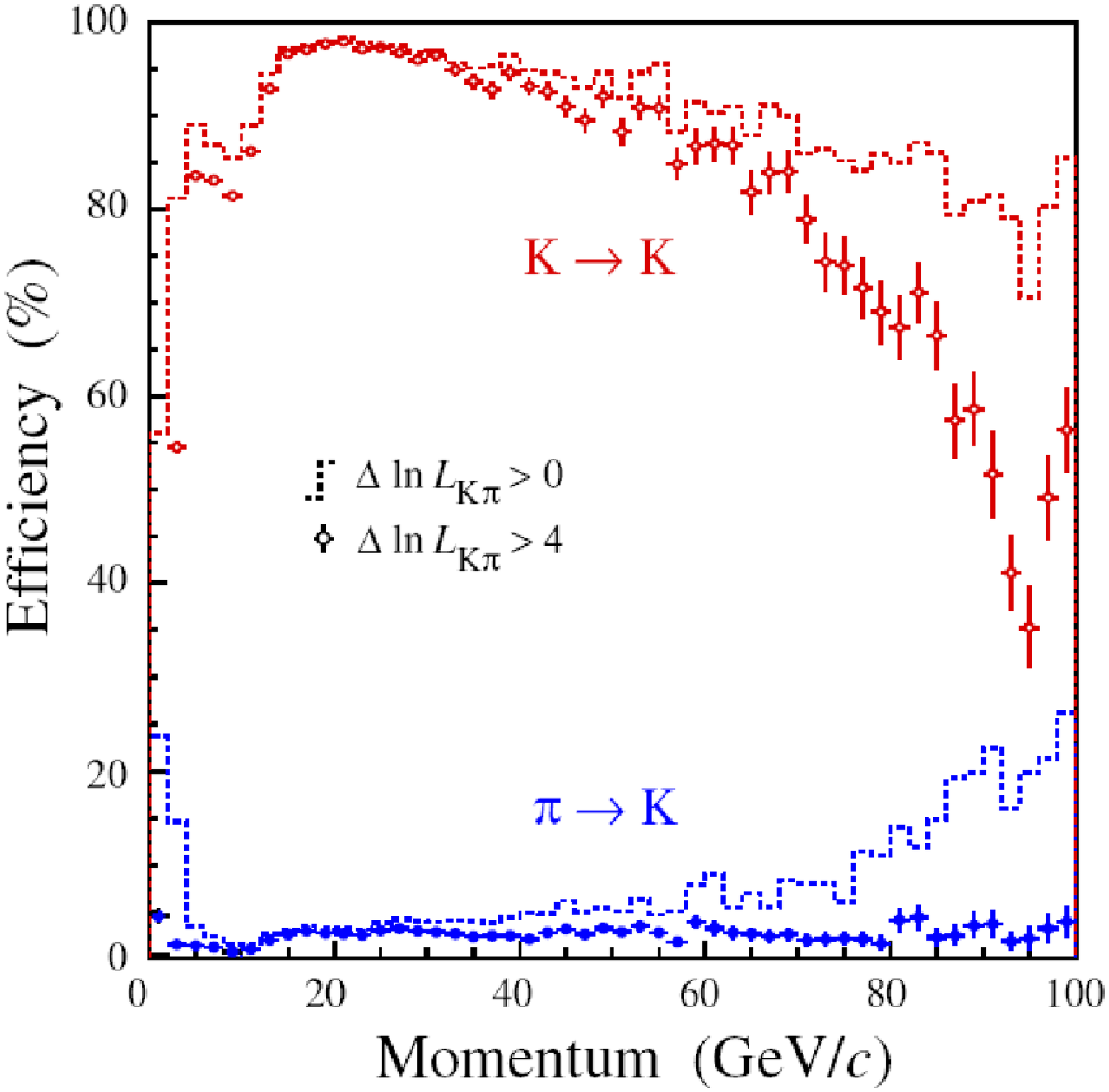}
\includegraphics[width=60mm]{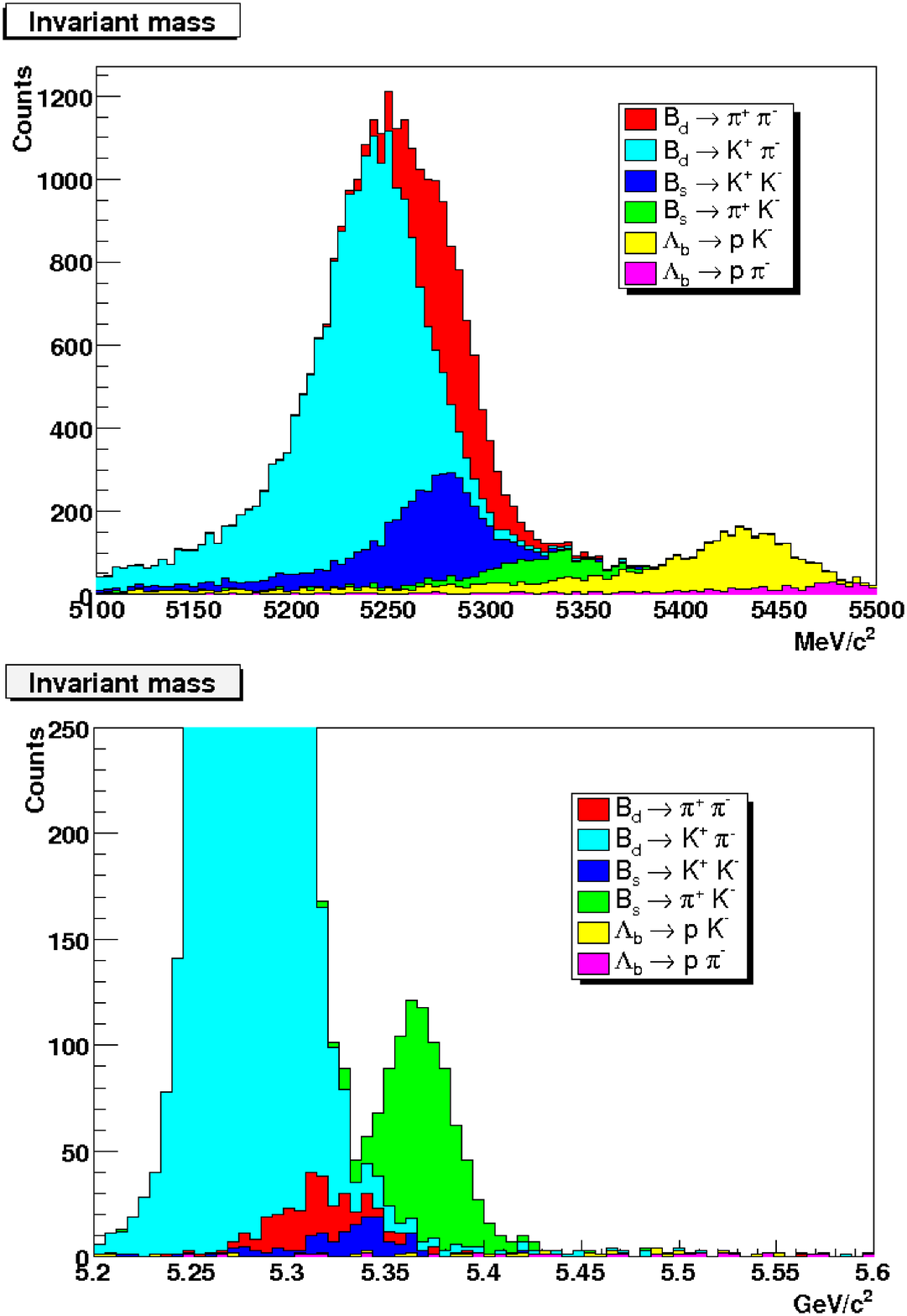}
\caption{The efficiency (above 30\%) and misidentification
(below 30\%) distributions
for two different delta log likelihood cuts, $\Delta$ln$L_{K\pi}>$0
as shown with the dashed line and $\Delta$ln$L_{K\pi}>$4 
shown with the open dots,
in the kinematic range 0 to 100 GeV/c. 
Right: B to hh mass spectra before (left) and after (right)
the use of particle ID information for the selection of the 
$B_s \to$ K $\pi$ decay.\label{fig:lhcbpid}} 
\end{figure*}

While the ATLAS and CMS detectors are concentrating all the efforts
on the analysis of B meson decays involving  muons, the LHCb 
experiment is able to identify also kaons and pions by means of two
Cerenkov detectors: the RICH1 and RICH2. 
The combined information coming from those two detectors is used to 
achieve a very good $\pi$,K 
separation in a wide kinematic range (2-100~GeV/c). The LHCb efficiency
and misidentification distributions are shown in the left 
plot of Figure
\ref{fig:lhcbpid} . These quantities will be calibrated with data
using pions and kaons from the 
D$^{*+} \to $D$^{0}$(K$^-\pi^+$)$\pi^+$ decay (expected rate: 16 Hz),
where a very pure sample of kaons
and pions can be selected using only kinematic cuts. 

The LHCb capability of identifying pions and kaons 
enhance the B physics program richness allowing the analysis
of non leptonic channels, like the B $\to$ hh or $B \to \phi\phi$
ones. These channels can be used
to improve the knowledge of the U spin symmetry breaking,
to constrain the CKM angle $\gamma$
(see par.~\ref{ckm:gamma}) or to measure the CKM
angle $\phi_s$ (see par.~\ref{ckm:phis})

The separation
of the several contributions ($\pi\pi$, K$\pi$, KK channels)
in the  B $\to$ hh analysis
can be achieved in LHCb only thanks to the excellent particle 
identification (PID) performance:
see for example the B$\to$hh mass spectra in 
the right up (down) plots of Figure \ref{fig:lhcbpid} 
obtained before (after) the use of PID information
for the selection of the $B_s \to$ K$\pi$ decay.

The muon identification in LHCb is provided by the muon system,
achieving a very high reconstruction efficiency 
($\epsilon >$ 95\%) 
with a low misidentification ($\leq$ 1\% for p$>$10 GeV/c).
These performances will be calibrated and measured on data
using dedicated samples. The muon PID can use the 
generic $\mu$ (50 Hz), prompt $J/\Psi$ (2 Hz) and 
$J/\Psi$ from B (0.03 Hz) events while the misidentification
can be measured using $\Lambda$ decays that provide a 
clean sample of protons and pions using only kinematic 
constraints and the hadrons reconstructed in the
B $\to$ hh decay chain (0.02 Hz).

\subsection{Tagging}
\label{sub:tag}

The tagging performances are determined using 
different strategies and algorithms exploiting the detector
capabilities in the different experiments.
These performances are expressed in terms of the
effective tagging efficiency ($\epsilon_{eff}$) defined as
$\epsilon_{eff} = \epsilon_{tag}(1-\omega)^2$, 
where $\epsilon_{tag}$ is the tagging efficiency 
(probability that the tagging procedure gives an answer)
 and $\omega$ is the wrong tag fraction 
(probability for the answer to be incorrect when a tag is present).

The ATLAS experiment quotes an $\epsilon_{eff}$ = 4.6\%
using the lepton and jet charge 
tagging algorithms on the opposite side candidates,
while the CMS experiment is still developing the tagging strategy
to assess the tagging performance.
The LHCb experiment is also using the
kaon tagging from the opposite side B and the pions ($B_d$)
or kaons ($B_s$) taggers from the same side as the signal candidate,
achieving an overall $\epsilon_{eff}$ = 6.6\%.
The tagging performances will be evaluated with data by using control
channels like B$^+ \to$ D$^0 \pi^+$.

\subsection{Proper Time Measurement}
\label{sub:pt}

The ability to measure the distance of flight of the B meson,
that is translated into the proper time using the
momentum information, is of crucial importance in CP analyses.
Since the $B_s$ mesons are oscillating faster than the $B_d$ ones
(CDF and D0 measure $\Delta$m$_s$ = 17.77~ps$^{-1}$ 
in good agreement with the expected SM value)
a good proper time resolution ($\sigma_t$)
is important in order to resolve
the oscillations.  The proper time resolution expected 
for the ATLAS (83~fs) and CMS (77~fs) experiments is
much larger than the one expected in LHCb (36~fs):
these values will be measured using the B mesons lifetimes
and, whenever possible, control channels.
For example in LHCb the $B_s \to D_s^+\pi^-$ decays
can be used to get a clean measurement of $\Delta$m$_s$
and of the $B_s$ oscillations, extracting the proper time resolution.

\subsection{B Physics Triggers}
\label{sec:trig}

In order to reduce the high rate of events coming from the pp collisions 
selecting only those that are likely to contain
a B meson, dedicated trigger algorithms are needed. 
The three LHC experiments are using a similar
approach: a first level trigger, hardware based, is used for
the fast selection of high p$_T$ events, while a second
level trigger, software based, is used to reduce the event rate to
an acceptable amount while keeping a maximum efficiency for 
signal events.

The ATLAS and CMS first level (L1) trigger has an output rate of 
$<$100~kHz and uses the information from the muon chambers
and the calorimeters to select high p$_T$ muons.
The LHCb first level (L0) trigger
has a 1~MHz output rate and uses the input from the muon system,
the calorimeters and the pileup system in order to select
h, $\mu$, e, $\gamma$ and $\pi^0$ over a minimum p$_T$ threshold.
The L0 trigger efficiency ($\epsilon_{L0}$) is $>$80\% for channels
with $J/\Psi$, while $\sim$40\% for other hadronic channels.

The software trigger for the LHC experiments is executed
after the full event readout. ATLAS and CMS experiments
have an overall output rate of 200~Hz: only $\sim$10\% of these
events can be used for B physics studies.
In addition, for the study of leptonic decays,
$O(1)$~kHz bandwidth of di-$\mu$ events is available. 

The LHCb High Level Trigger (HLT) has an output rate of 2~kHz.
Starting from the L0 confirmation the
trigger bandwidth is divided between inclusive and 
exclusive selections as follows: 200~Hz is the 
bandwidth allocated for the core B physics program
(exclusive reconstruction),  600~Hz are devoted to 
the reconstruction of high mass dimuons events (allowing
for an unbiased proper time selection of $J/\Psi$ decays),
900~Hz are available for an inclusive B meson selection
using semileptonic decays ($B \to \mu$) mainly devoted
to data mining studies and the remaining 300~Hz will
be used for the $D^*$ reconstruction to be used for the particle
ID calibration and charm studies.

\section{B PHYSICS PROGRAM}
\label{sec:bprog}

The B physics program of the ATLAS, CMS and LHCb experiments
covers several different areas: from the cross section
to the leptonic decays studies, from the rare decays to the 
CP violation studies, from the NP searches to the search
of new particles trough the Dalitz analysis and data mining.

All the studies reported here have been performed on events
coming from the standard simulation, digitization and 
reconstruction experiments software. 
The statistics used to quote the results,
unless specified otherwise, do refer to one fourth of a
nominal year running: 2.5 fb$^{-1}$ for ATLAS and CMS and 
0.5 fb$^{-1}$ for LHCb. These data samples will be likely available 
in 2009. 

Only few topics have been chosen and will be reported here:
the study of b $\to$ s transitions 
in the $B_s \to J/\Psi \phi$ channel, b$\to$sll decays,
the measurement of the $B_s \to \mu\mu$ branching ratio, the measurement
of the CKM angle $\gamma$ and of the B production cross sections.

\subsection{$B_s \to$ J/$\Psi \phi$}
\label{ckm:phis}

The study of the $B_s \to J/\Psi \phi$ decays can be used
to measure the CKM angle $\phi_s$,
predicted to be very small by the SM:
\begin{equation}
\phi_s [SM] = –arg(V_{ts}^2) = –2\lambda^2\eta = –0.0368 \pm 0.0018 
\end{equation}
and hence sensible to 
NP processes that could give measurable contributions.

From the experimental point of view this decay is particularly 
suited for the detection and reconstruction in all the LHC experiments,
since it involves 2 muons from the $J/\Psi$ decay,
allowing an efficient trigger,
and has an high predicted BR: $\sim$3$\cdot$10$^{-5}$.

\begin{figure*}[t]
\centering
\includegraphics[width=120mm]{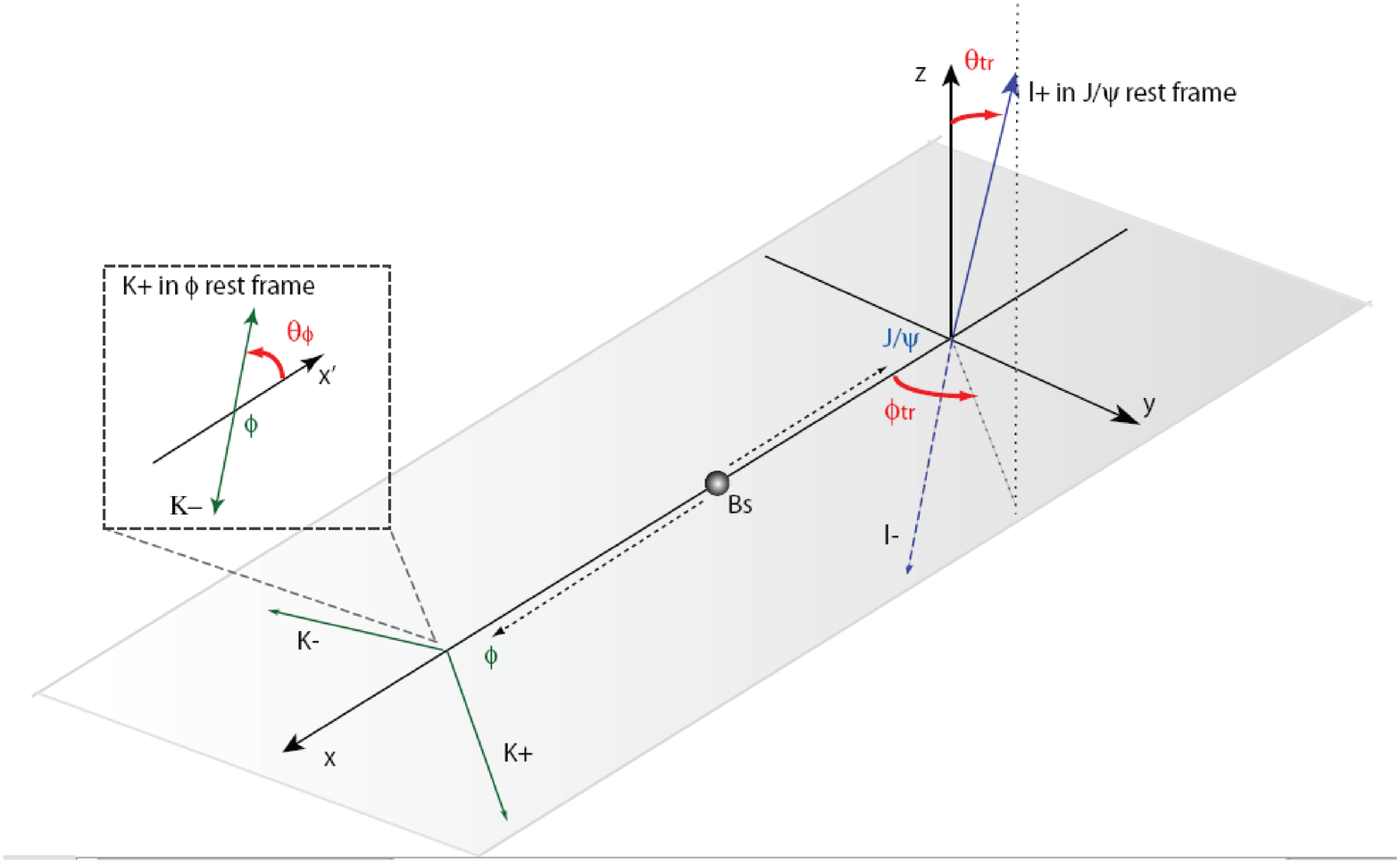}
\caption{Definition of angles $\theta_\phi$, $\theta_{tr}$ and 
$\phi_{tr}$, used in the three angle analyis of the 
$B_s \to J/\Psi \phi$ decay.\label{fig:anglesJpsiPhi}} 
\end{figure*}

The value of $\phi_s$(-2$\beta_s$) can be extracted from the measurement of 
the time dependent CP asymmetries A$_{CP}$(t):
\begin{equation}
A_{CP}(t) = \frac{-\eta_f sin \beta_s sin(\Delta m_st)}{cosh(\Delta\Gamma_st/2)-\eta_fcos\beta_ssinh(\Delta\Gamma_st/2)}
\end{equation}
where $\eta_f$ = 1(-1) for odd(even) CP states.
Since the decay final state is not a pure CP eigenstate,
an angular analysis is needed that involves the
three angles $\theta_\phi$, $\theta_{tr}$, $\phi_{tr}$
defined as shown in Figure \ref{fig:anglesJpsiPhi}. 
Together with the angles, the measurement
of the $B_s$ candidates mass, proper time and tagging informations
are needed as input to the analysis.

The sensitivity to the $\phi_s$ angle heavily depends 
on the tagging performance
(entering trough the $\epsilon_{eff}$ term) and the proper time 
resolution: the different detectors performances have been already 
reported in paragraphs \ref{sub:tag} and \ref{sub:pt}.

The specific cuts, used in the selections of the signal events,
are described in~\cite{bib:jpsiphisel}. 
The number of reconstructed events is shown in the first
row of Table \ref{tab:jpsiphi}, while the number of the 
flavour tagged events is given in the second row.

The candidates have mass resolutions of $\sim$15~MeV/c$^2$ with
background over signal ratios (B/S) in the range 0.25$-$0.12
(see the third and fourth rows in Table \ref{tab:jpsiphi}).
It should be noted that the LHCb result is obtained without using
the $J/\Psi$ mass constraint allowing a better background control.

\begin{table}[t]
\begin{center}
\caption{B mass resolutions, B/S ratios and number of selected events
for the analysis of $B_s \to J/\Psi \phi$ events in the LHC experiments.}
\begin{tabular}{|l|c|c|c|}
\hline
\textbf{Parameter} & \textbf{ATLAS} & \textbf{CMS} & \textbf{LHCb} \\ \hline
N$_{rec}$ & 23k & 27k & 33k \\
N$_{rec}^{eff-tag}$ & 1 k & - & 2.2 k \\ 
$\sigma_m$[MeV/c$^2$] & 16.5 & 14 & 14 \\
B/S & 0.18 & 0.25 & 0.12 \\
$\sigma(\phi_s)$ & 0.159 & - & 0.042 \\
$\sigma(\Delta\Gamma_s)/\Delta\Gamma_s$ & 0.41 & 0.13 & 0.12 \\
\hline
\end{tabular}
\label{tab:jpsiphi}
\end{center}
\end{table}

The resolutions on $\phi_s$ and $\Delta\Gamma_s$,
that can be achieved with data collected during 2009
are shown in the last two columns of Table  \ref{tab:jpsiphi}.

With the quoted values, LHCb is able to 
exclude/measure NP contributions to $\phi_s$
beyond the SM (see for example \cite{bib:ligeti}
for a discussion of the impact of LHCb measurement of $\phi_s$
in constraining NP contributions to b$\to$s transitions). 
Further improvements could also come by adding 
other decay channels,
like $J/\Psi \eta$, $\eta_c \phi$ or $D_s^+ D_s^-$,
to the $\phi_s$ analysis.
By 2013, ATLAS and CMS are expected to achieve the 
SM level sensitivity, while LHCb will test the SM prediction
at the 5$\sigma$ level.

\subsection{$B_s \to \mu\mu$}

The $B_s \to \mu\mu$ decay is predicted in the SM to be 
very rare, since it involves flavour changing neutral currents
and experiences a large helicity suppression 
($\sim$m$_\mu$/m$_b$), with a BR at the level 
of 10$^{-9}$. 

\begin{figure*}[t]
\centering
\includegraphics[width=70mm]{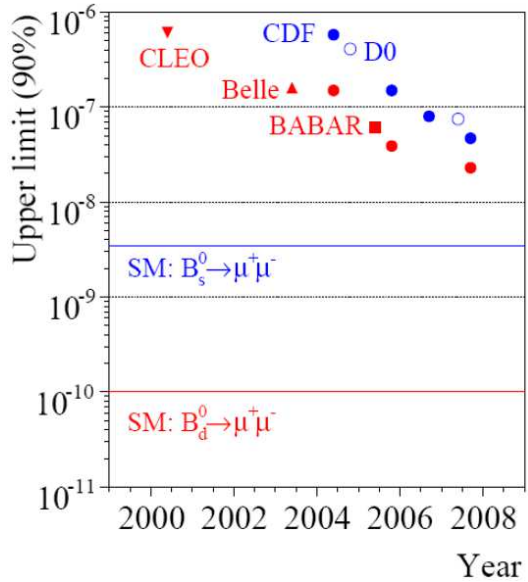}
\caption{Current experimental overview of
$B_s \to \mu\mu$ (blue/darker markers) and $B_d \to \mu\mu$ 
(red/lighter markers)  BR measurements, as a function of time.
\label{fig:bstomumu}} 
\end{figure*}

Various attempts have already been made to measure
the BR: the current experimental situation, concerning both
$B_d$ and $B_s$ decays, is summarized in Figure \ref{fig:bstomumu}
where the current best limits are coming from 
CDF~\cite{bib:bsmuCDF} (BR $<$ 4.7 10$^{-8}$ $@$90\% CL) and 
D0~\cite{Scuri:2007py} (BR $<$ 7.5 10$^{-8}$ $@$90\% CL) 
measurements.

The study of this decay is of particular interest when
probing NP models, like the MSSM, that allows for large 
modifications of the BR: the contribution
of a tan$^6\beta$/M$^4_A$ term, for example, can significantly
enhance the measured value if the 
tan$\beta$ value is large, as 
suggested by other experimental results
like b$\to$s$\gamma$,
(g-2)$_\mu$ and $B \to \tau\nu$.

The measurement of
the $B_s \to \mu\mu$ BR can thus play a major role
in constraining/measuring the NP contributions and 
even help selecting/ruling out a given theoretical model
\cite{bib:parisi}.

The analysis strategies pursued are different in LHCb~\cite{bib:bsmumulhcb}
and in ATLAS and CMS~\cite{bib:rareAC}. 
In the LHCb approach the topological informations
are used to build a geometrical likelihood for each event (GL). Then
the three dimensional space built with the 
GL, mass and particle ID informations 
is divided for all the events in N bins.
For each bin the expected number of events for the signal 
and for the signal plus background hypothesis are computed and 
used to place a limit (or evaluate the measurement
sensitivity) on the BR. 

The ATLAS and CMS experiments are
instead using a cut and counting approach.
The geometrical and particle ID
informations are used to select the events by applying cuts.
The mass distribution is then used to count the events:
CMS opens a $\pm$2.3 $\sigma$ window, while
ATLAS performs a bayesian number of events estimate.

\begin{table}[t]
\begin{center}
\caption{Expected B meson mass resolutions and number of signal 
and background events
obtained in the study of $B_s \to \mu\mu$ decays 
for the different LHC experiments.}
\begin{tabular}{|l|c|c|c|}
\hline
\textbf{Parameter} & \textbf{ATLAS} & \textbf{CMS} & \textbf{LHCb} \\ \hline
$\sigma_m$[MeV/c$^2$] & 67 & 43 & 20 \\
N$_{sig}$ & 7 & 6 & 30 \\
N$_{bkg}$ & 20 & 14 & 83 \\ 
\hline
\end{tabular}
\label{tab:bsmumu}
\end{center}
\end{table}

The normalization channel used in the BR measurement is 
the B$^+ \to J/\Psi K^+$: one million of such
events is expected for each fb$^{-1}$ of data in all the LHC experiments.
The LHCb experiment also foresees the use of B$\to$hh decays
as a control channel, while the background will be 
extracted from the side bands.

\begin{figure*}[t]
\centering
\includegraphics[width=70mm]{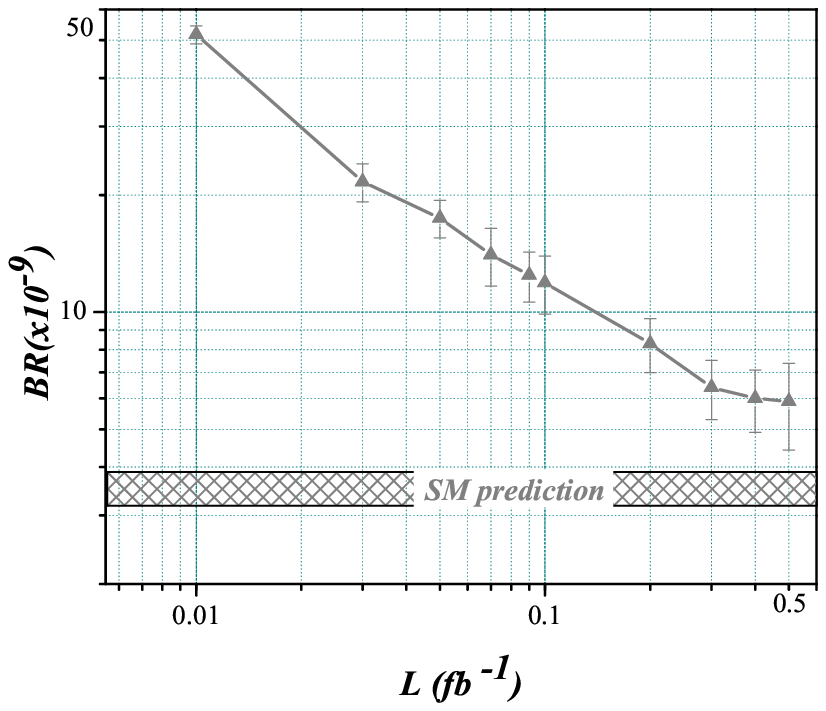}
\includegraphics[width=70mm]{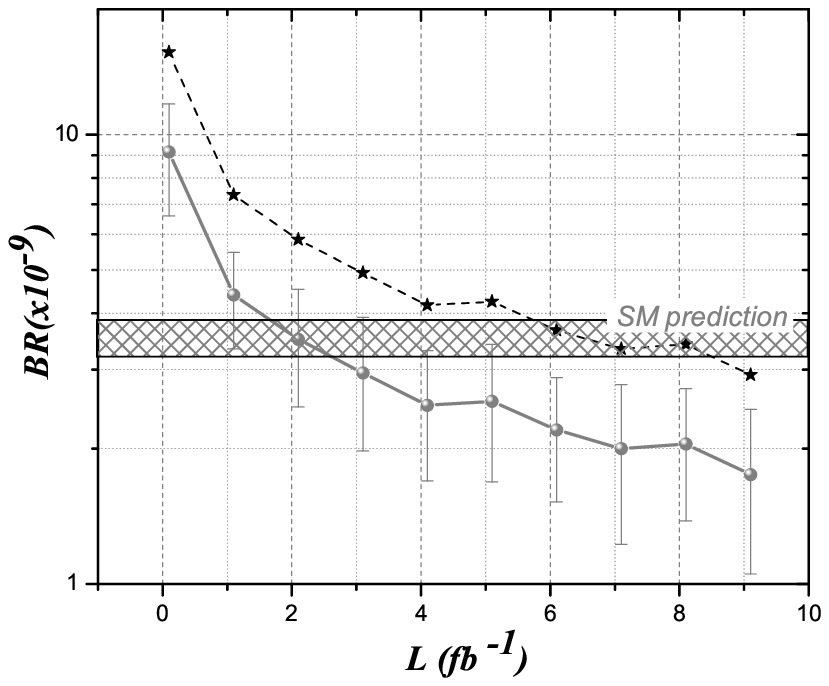}
\caption{Left: BR($B_s^0 \to \mu^+\mu-$) excluded at 90\% CL as a 
function of the integrated luminosity (L) if no signal is present. 
Right: Luminosity needed for the observation of a given BR at 3$\sigma$
(grey circles) and 5$\sigma$ (black stars) level.
\label{fig:bsmumuL}} 
\end{figure*}

\begin{figure*}[b]
\centering
\includegraphics[width=70mm]{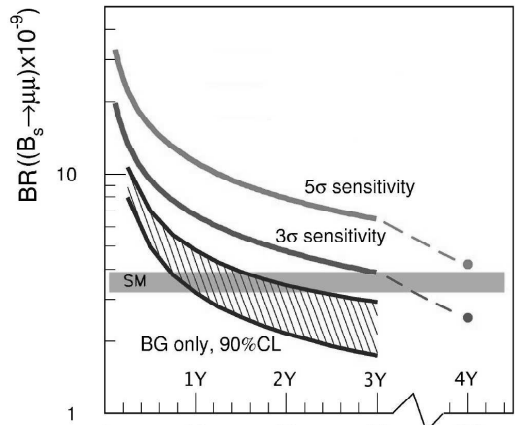}
\caption{ATLAS perspective for measuring the
$B^0_s \to \mu^+\mu^-$
branching ratio as a function of the time (integrated luminosity).
\label{fig:bsmumuA}} 
\end{figure*}

The expected number of events after a nominal year of running
(2 fb$^{-1}$ LHCb, 10 fb$^{-1}$ ATLAS and CMS) is quoted in the second
and third rows of Table \ref{tab:bsmumu}. 
The sensitivities of LHCb as a function of the integrated luminosity
are given in Figure \ref{fig:bsmumuL}, while the ATLAS
result is shown in Figure \ref{fig:bsmumuA}.

The LHCb experiment has the potential of measuring a BR
of 9(15)$\cdot$10$^{-9}$ at 3(5)$\sigma$ level with
0.1 fb$^{-1}$ of data and 5(9)$\cdot$10$^{-9}$ with 0.5 fb$^{-1}$.
The SM BR can be assessed with a 3(5)$\sigma$ evidence(observation)
with 2(6)fb$^{-1}$.

The ATLAS and CMS experiments, when properly rescaling 
the integrated luminosity, are contributing with comparable
performances.

\subsection{b$\to$sll decays}

The study of exclusive b$\to$sll decays is another way
to look for NP effects in b$\to$s transitions at an hadron collider:
the leptons in the decay provide an efficient trigger
while the exclusive reconstruction reduce the background contribution.

These semileptonic decays proceed trough
suppressed loops in SM and thus are sensitive to NP
contributions, that can be of SM size, affecting
the BR and the angular distributions.

Accurate SM predictions can be obtained for some
specific observables, like the Forward-Backward asymmetry (A$_{FB}$),
the invariant $\mu\mu$ mass ($q^2$) value where $A_{FB}$ reach its
zero value (s$_0$), the transversal asymmetries and the 
ratio of BR for ee and $\mu\mu$ decay modes, where it is possible
to obtain 

As an example, for the B$\to K ^*$ll decay channel,
the SM predictions \cite{bib:bsll} are:
BR($B_d\to K^*\mu\mu$)=(1.22$^{+0.38}_{-0.32}$)10$^{-6}$
and zero crossing of A$_{FB}$ 
\begin{equation}
s_0 = s_0(C_7,C_9) = 4.39^{+0.38}_{-0.35} GeV^2
\end{equation}

In the LHCb experiment a selection yielding a B$_{bb}$/S
of 0.2$\pm$0.1 has been developed: 1.8~k events are expected
in 0.5~fb$^{-1}$, to be compared with the 450 events expected
from the B factories with a 2~ab$^{-1}$ integrated luminosity.
With 2 fb$^{-1}$ the sensitivity on s$_0$ is 0.46 GeV$^2$, while
with 10 fb$^{-1}$ the present theoretical precision (0.27 GeV$^2$)
is reached.

The ATLAS experiment expect $\sim$3~k events with 30~fb$^{-1}$, achieving
a $\sim$4\% statistical error on the A$_{FB}$ distribution in
q$^2$ bins. The CMS studies have just started.

\subsection{CKM $\gamma$ angle}
\label{ckm:gamma}

The measurement of the CKM angle $\gamma$ trough the analysis
of hadronic decays, like B$\to$DX or B$\to$hh, is only
feasible and addressed so far in the LHCb experiment,
the reason being the need of kaon and pion identification that
is not available in ATLAS and CMS.

The selection of several decay modes has been
studied in detail: the modes under study,
together with their expected number of events
(in 2~fb$^{-1}$ integrated luminosity) and inclusive \bbar background
over signal (B$_{bb}$/S) ratios
are given in Table \ref{tab:btodsel}.

\begin{table}[t]
\begin{center}
\caption{Expected event yields and inclusive \bbar over signal events ratio
for several B to D decays studied in LHCb.}
\begin{tabular}{|l|c|c|}
\hline
\textbf{Decay mode} & \textbf{Event yield} & \textbf{B$_{bb}$/S} \\ \hline
B$^{-,+}\to$D(K$\pi$)K$^{-,+}$ favoured & 28k & 0.6 \\
B$^{-,+}\to$D(K$\pi\pi\pi$)K$^{-,+}$ favoured & 28k & 0.6 \\
B$^{-,+}\to$D(K$\pi$)K$^{-,+}$ supp. & 100 & $>$2 \\
B$^{-,+}\to$D(K$\pi\pi\pi$)K$^{-,+}$ supp. & 200 & $>$2 \\
B$^{-,+}\to$D(hh)K$^{-,+}$ & 4k & 2 \\
B$^{-,+}_s\to$D$_s$K$^{-,+}$ & 6.2k & 0.2 \\
\hline
\end{tabular}
\label{tab:btodsel}
\end{center}
\end{table}

Several strategies can be pursued in 
order to measure $\gamma$. The analysis of B$\to$DK channels,
through the ADS or GLW and GGSZ approaches~\cite{gamma}
uses the large statistics
that can be collected at the cost of a dependence on the
D strong phases affecting the sensitivity. 
The study of the less abundant $B_s \to$D$_s$K
decay mode yields instead a clean extraction of $\gamma$
from the interference of b$\to$u and b$\to$c transitions
in the $B_s$ mixing. Other Dalitz and 4 body decay analyses
can be pursued as well, yielding the expected resolutions
quoted in Table \ref{tab:btodpres}

\begin{table}[t]
\begin{center}
\caption{Expected precision on the CKM angle $\gamma$
for several B to D decays studied in LHCb. The B and D decay modes and
the analysis method are shown for each decay in the first three columns.}
\begin{tabular}{|l|c|c|c|}
\hline
\textbf{B Decay mode} & \textbf{D decay mode} & \textbf{Method} & \textbf{$\sigma(\gamma)$} \\ \hline
$B_s\to$D$_s$K & KK$\pi$ & tagged, A(t) & 10$^{\circ}$\\
B$^+\to$DK$^+$ & K$\pi$+K3$\pi$+KK/$\pi\pi$ & counting, ADS,GLW & 5$-$13$^{\circ}$\\
B$^+\to$D$^*$K$^+$ & K$\pi$ & counting, ADS,GLW & Under study\\
B$^+\to$DK$^+$ & K$_s\pi\pi$ & Dalitz, GGSZ & 7$-$12$^{\circ}$\\
B$^+\to$DK$^+$ & KK$\pi\pi$ & 4 body Dalitz, GGSZ & 18$^{\circ}$\\
B$^+\to$DK$^+$ & K$\pi\pi\pi$ & 4 body Dalitz, GGSZ & Under study\\
B$^0\to$DK$^{*0}$ & K$\pi$+KK+$\pi\pi$ & counting, ADS+GLW & 9$^{\circ}$\\
\hline
\end{tabular}
\label{tab:btodpres}
\end{center}
\end{table}

\subsection{Cross section for \bbar production}

The current prediction of $\sigma$(\bbar) $\sim$ 500~$\mu$b 
comes from the extrapolations of Tevatron results:
a precise measurement of such cross section is of major importance
in order to test the MC simulations, the NLO QCD calculations 
used for the extrapolations and the Parton Density Functions (PDF) knowledge.
For the rare decays, NP and new particle searches the knowledge
of $\sigma$(\bbar) is also important to achieve a good estimate
of the background level.

Inclusive and exclusive strategies can be used to measure 
the cross section. All the LHC experiments plan to use 
J/$\Psi$ events to measure the production rate:
the muons in the decay ensure an high efficiency
in triggering and reconstructing those channels.
In ATLAS and CMS experiments an additional requirement
on the transverse energy (E$_T$) 
is used in order to enhance the b quark component in
the data sample.

The statistical error that can be achieved is
$\sim$ 1\%, already with small data samples 
(L $\sim O(10) pb^{-1}$). 
More detailed analyses, aiming at a full p$_T$ scan, achieving 
a resolution of $\sim$10\% in all bins,
have been performed in CMS assuming L $\sim$ 10~fb$^{-1}$.

\section{CONCLUSIONS}

The LHC program is proceeding without major delays:
first beam is expected in September and the first collisions soon afterwards.
There are currently three 
experiments getting ready for the B physics challenge: with the
expected performance they will be able to test SM 
and beyond SM (NP) effects.

The first analysis approach reported here is related to the
b$\to$s observables, that can cleanly reveal some NP effects.
A first goal of the LHC experiments is the measurement of the 
$B_s \to\mu\mu$ BR down to its current SM prediction and of the 
$\phi_s$ angle with a 0.04 absolute precision. These goals can be
achieved with data collected during 2009.

The analysis of b$\to$sll decays is also presented: 
with few years of data taking the measurements
precision will reach the level of
present theoretical uncertainty. 

The LHCb detector will also address
the study of B$\to$DK and B$\to$hh decay modes that
can be used to reduce the CKM $\gamma$ angle uncertainty down to 
10$^{\circ}$, with data collected during 2009.

The measurement of \bbar production
cross section within few \%,  with early data, and 
the full p$_T$ scan ($@$ 10\% level), 
with few years of data taking, can
also be achieved.

The LHC experiments, whose B physics program
highlights have been reported here, are
ready to provide new measurements that will
enrich the LHC program by exploring a new whole experimental
region in the b quark phenomenology.

\begin{acknowledgments}
The author wish to thank A. Policicchio (ATLAS)
and U. Langenegger (CMS) for their help in preparing
these proceedings.
\end{acknowledgments}


\begin{thebibliography}{9}   

\bibitem{lhc:det}
L.~Evans and P.~Bryant,  ``LHC Machine'',
2008, JINST, 3 S08001, available at http://www.iop.org/EJ/journal/1748-0221  

\bibitem{lhcb:det}
The LHCb Collaboration,  ``The LHCb Detector at LHC'',
2008, JINST, 3 S08005, available at               
http://www.iop.org/EJ/journal/1748-0221  

\bibitem{atlas:det}
The ATLAS Collaboration, ``The ATLAS Experiment at the CERN Large Hadron Collider'',
2008 JINST 3 S08003, available at               
http://www.iop.org/EJ/journal/1748-0221  

\bibitem{cms:det}
The CMS Collaboration, ``The CMS experiment at the CERN LHC'',
2008 JINST 3 S08004, available at               
http://www.iop.org/EJ/journal/1748-0221  

\bibitem{tev:det} CDF Collaboration, D.Acosta {\it et al, Phys. Rev.} D 
{\bf 71}, 032001 (2005), D0 Collaboration, V.M.Abazov {\it et al, Nucl. 
Instrum. Methods} {\bf A} 565, 463 (2006).

\bibitem{tev:dms} A. Abulencia et al., Phys. Rev. Lett. 97, 242003 (2006)

\bibitem{tev:bs} The CDF Collaboration, ``First Flavor Tagged Determination of Bounds on Mixing Induced CP Violation in $B_s \to$ J/psi phi Decays'', http://arxiv.org/abs/0712.2397v1. The D0 Collaboration, ``Measurement of $B_s^0$ mixing parameters from the flavor-tagged decay'', http://arxiv.org/abs/0802.2255.

\bibitem{bib:jpsiphisel} P. Clarke, C. McLean, A. Osorio-Oliveros, ``Sensitivity studies to $\beta_s$ and $\Delta\Gamma_s$ using the full $B_s \to J\Psi \phi$ angular analysis at the LHCb'', CERN-LHCB-2007-101

\bibitem{bib:ligeti} Z. Ligeti {\it et al.}, ``Implications of the measurement of the $B^0_s-\bar B^0_s$ mass difference'' , Phys. Rev. Lett. 97, 101801 (2006)


\bibitem{bib:bsmuCDF} The CDF Collaboration, ``Search for $B_s \to \mu^+\mu^-$ and $B_d \to \mu^+\mu^-$ Decays with 2fb$^{-1}$ of ppbar Collisions'', http://arxiv.org/abs/0712.1708v1

\bibitem{Scuri:2007py} F.~Scuri  [CDF Collaboration and D0 Collaborations], ``Measurements of $B$ rare decays at the Tevatron,'' arXiv:0705.3004 [hep-ex].

\bibitem{bib:parisi} G.~Isidori e P.~Paradisi, Phys Lett. B {\bf 639}, 499 (2006)

\bibitem{bib:bsmumulhcb} D. Martinez Santos, ``$B^0_s \to \mu^+\mu^-$ in LHCb'',  CERN-LHCb-2008-018

\bibitem{bib:rareAC} A. Policicchio and G. Crosetti, ``Study of DiMuon Rare Beauty Decays with ATLAS and CMS'', Eur. Phys. J. C {\bf 55}, 173-176 (2008)
DOI: 10.1140/epjc/s10052-008-0594-6

\bibitem{bib:bsll} M.~Beneke {\it et al.}, ``Exclusive radiative and electroweak $b \to d$ and $b \to s$ penguin decays at NLO'', arXiv:0412400 [hep-ph]

\bibitem{gamma} D.~Atwood, I.~Dunietz, A.~Soni, Phys. Rev. Lett. {\bf 78}, 3257 (1997) and and Phys. Rev. D {\bf 63}, 036005 (2001); M.~Gronau and D.~Wyler, Phys. Lett. B {\bf 265}, 172 (1991); M.~Gronau and D.~London, Phys. Lett. B {\bf 253}, 483 (1991); A.~Giri, Yu.~Grossman, A.~Soffer and J.~Zupan, Phys. Rev. D {\bf 68}, 054018 (2003).

\end{thebibliography}
\end{document}